\documentclass{mn2e}
\newif\ifAMStwofonts
\AMStwofontstrue
\usepackage{graphicx}
\newcommand{\etal}{et~al.~}

\title{X-ray Emission from Haloes of Simulated Disc Galaxies}
\author[S. Toft et al.]
       {S. Toft,$^1$\thanks{E-mail: toft@astro.ku.dk},
J. Rasmussen$^1$, J. Sommer-Larsen$^2$ and  K. Pedersen,$^1$\\
 $^1$Astronomical Observatory, Copenhagen University, Juliane Maries Vej 30, DK-2100 Copenhagen \O, Denmark \\ $^2$Theoretical Astrophysics Center, Juliane Mariesvej 30, DK-2100 Copenhagen \O, Denmark  }
\date{}

\pagerange{\pageref{firstpage}--\pageref{lastpage}}
\pubyear{2001}

\begin{document}

\maketitle

\label{firstpage}

\begin{abstract}
Bolometric and 0.2-2 keV X-ray luminosities of the hot gas haloes of simulated
disc galaxies have been calculated at redshift $z$=0. The TreeSPH simulations are
fully cosmological and the sample of 44 disc galaxies span a range in
characteristic circular speeds of $V_c$ = 130-325 km s$^{-1}$. The galaxies have been
obtained in simulations
with a considerable range of physical parameters, varying the baryonic
fraction, the gas metallicity, the meta-galactic UV field, the cosmology, the
dark matter type, and also the numerical resolution. The models are found to
be in agreement with the (few) relevant X-ray observations available
at present. The amount of hot gas in the haloes is also consistent
with constraints from pulsar dispersion measures in the Milky Way.
Forthcoming XMM and Chandra observations should enable much more stringent
tests and provide constraints on the physical parameters.
We find that simple cooling flow models over-predict X-ray luminosities
by up to two orders of magnitude for high (but still realistic)
cooling efficiencies relative to the models presented here.
Our results display a clear trend that {\it increasing} cooling efficiency
leads to {\it decreasing} X-ray luminosities at $z$=0. The reason is found
to be that increased cooling efficiency leads to a decreased fraction of
hot gas relative to total baryonic mass inside of the virial radius at 
present. At gas metal
abundances of a third solar this hot gas fraction becomes as low as
just a few
percent. 
We also find that most of the X-ray emission comes from the inner
 parts ($r\la 20$ kpc) of the hot galactic haloes.
Finally, we find for realistic choices of the physical parameters
that disc galaxy haloes possibly were {\it more than one order of magnitude}
brighter in soft X-ray emission at $z$$\sim$1, than at present.

\end{abstract}

\begin{keywords}
methods: N-body simulations  --cooling flows --galaxies: evolution --galaxies: formation --galaxies: halos --galaxies: spiral --X-rays: galaxies

\end{keywords}

\section{Introduction}
In disc galaxy formation models infall of halo gas onto the disc due
to cooling is a generic feature. However, the gas accretion rate and 
hot gas cooling 
history are at best uncertain in all models so far. It is thus not clear to 
which extent 
the gas cooling out from the galaxy's halo is replenishing that which is
consumed by star formation in the disc.
Such continuous gas infall is essential to
explain the extended star formation histories of isolated spiral
galaxies like the Milky-Way and the most likely explanation of the ``G-dwarf
problem'' --- see, e.g., Rocha-Pinto \& Marciel (1996) and Pagel (1997).

At the virial temperatures of disc galaxy haloes the dominant cooling
mechanism is thermal bremsstrahlung plus atomic line emission.
The emissivity, increasing strongly with halo gas density, is expected to 
peak fairly close to the disc and decrease outwards, 
and if the cooling rate is significant
the X-ray flux may be visible well beyond the optical radius of a galaxy.

Recently, Benson \etal (2000) compared ROSAT observations 
of three
massive, nearby and highly inclined disc galaxies with predictions of simple
cooling flow models of galaxy formation and evolution.  
They showed that these models predict about {\emph{an order of
magnitude}} more X-ray emission from the galaxy haloes (specifically
from a 5-18 arcmin annulus around the galaxies) than observational detections and upper limits.

In this paper we present global X-ray properties of the haloes of a large, 
novel sample of model disc galaxies at redshift $z$=0. The galaxies result from
physically realistic gravity/hydro simulations of disc galaxy 
formation and evolution in a cosmological context. We find that our model 
predictions of X-ray properties
of disc galaxy haloes are consistent with observational detections and
 upper limits. Given the  
results of the theoretical models of
Benson \etal we list the most important reasons why simple cooling flow models
over-predict the present day X-ray emission of disc galaxy haloes.

In section 2 we give a very short description of the disc galaxy simulations.
In section 3 we briefly describe the X-ray halo emission calculations and in section 4 the
results obtained. Section 5 constitutes the discussion and section 6
the conclusion.
\begin{figure}
\resizebox{\hsize}{!}{{\includegraphics{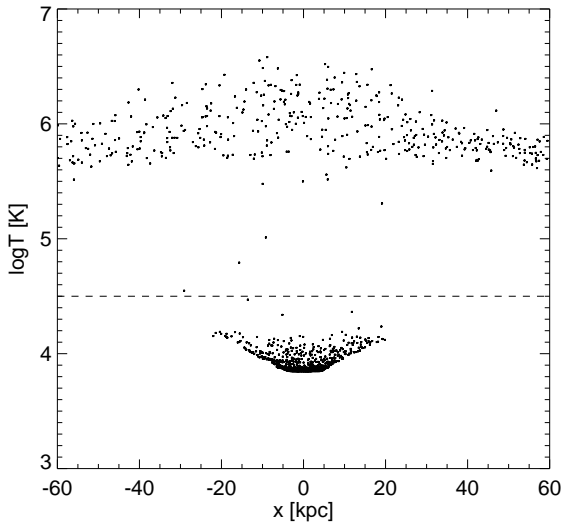}}}
\caption{The figure shows the temperature of the SPH gas
particles in a typical disc galaxy from our simulations versus their x
coordinate (one of the axes in the disc).  The ``cold'' (log$T<4.5$) gas which
is primarily situated
in the disc is removed from the catalogues since it does not
contribute to the X-ray flux. }
\label{figT}
\end{figure}  
 
\section{Disc galaxy simulations}
We have in recent years developed novel models of formation and evolution of
disc galaxies. The model disc galaxies result from {\it{ab initio}},
fully cosmological ($\Omega_M=1$ or $\Omega_M+\Omega_{\Lambda}=1$),
gravity/hydro simulations. These simulations are started at a
sufficiently high redshift ($z_i$=20-40) that the density
perturbations are still linear and are then evolved through the entire
non-linear galaxy formation regime to the present epoch ($z$=0). The
code uses a gridless, fully Lagrangian, 3-D TreeSPH code incorporating
the effects of radiative cooling and heating (including the effects of
a meta-galactic UV field), inverse Compton cooling, star formation, and 
energetic stellar feedback processes.

A major obstacle in forming realistically sized disc galaxies
in such simulations is the so-called ``angular momentum
problem'' (e.g., Navarro \& White 1994, Sommer-Larsen et al. 1999). 
We overcome this problem in two different ways: a) Using
cold dark matter (CDM) + stellar feedback processes
\cite{sommerlarsen02} or b) Using warm dark matter (WDM)
\cite{sommerlarsendolgov}. A total of 44 such disc galaxy
models with characteristic circular velocities in the range  
$V_c$ = 130--325 ${\rm km\,s^{-1}}$ form the basis of the predictions 
presented in this paper. 
The simulations initially consist of 30000-400000 SPH+DM particles and
in the majority of them some of the SPH particles are turned into star
particles over the course of the simulation --- for details
about the TreeSPH simulations we refer the reader to the above quoted 
references.

Most of the simulations were run with primordial gas composition (76\% H and
24\% He by mass) under the
assumption that the inflowing, hot gas is fairly unenriched in heavy elements.
To test the effects of metal abundance eight $\Lambda$CDM simulations 
(four with ``universal'' baryonic fraction $f_b$=0.05 and four with $f_b$=0.10)
were run with a gas abundance of 1/3 solar (specifically [Fe/H]=-0.5). This is
the metal abundance of the intracluster medium and can probably be considered
a reasonable upper limit to the metal abundance of the hot gas in disc galaxy
haloes.

\begin{figure}
\resizebox{\hsize}{!}{{\includegraphics{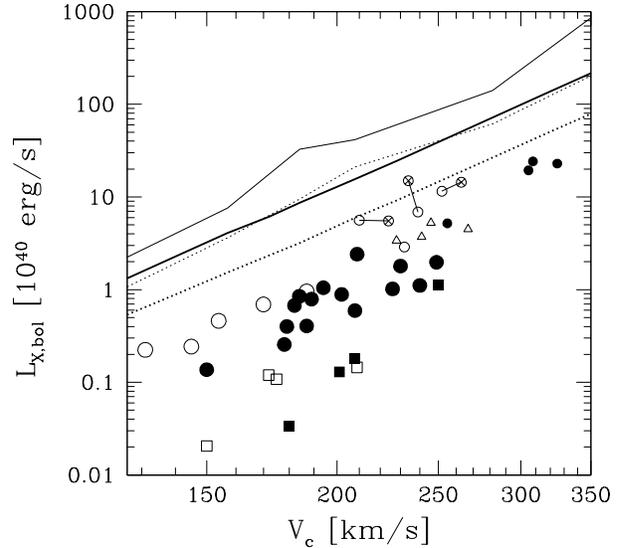}}}
\caption{Bolometric luminosity as a function of characteristic circular
speed.
{\bf Small symbols}: Flat $\Omega_M=1.0$ cosmology:
Open symbols: baryon fraction $f_b$=0.05, filled circles
$f_b$=0.1. Triangles: without UV field, non-triangles: with a UV field
of the Efstathiou (1992) type. 
Connected symbols are the same galaxies run with medium (open
circles) and high (open circles with crosses) resolution. All  
simulations represented by small symbols have primordial abundance.
{\bf Large symbols}: Flat ($\Omega_{\Lambda},\Omega_M)=(0.7,0.3)$ cosmology: 
Open symbols: $f_b$=0.05, filled symbols $f_b$=0.1. Circles correspond to
primordial abundance and with a Haardt \& Madau (1996) UV field, squares
correspond to $Z=1/3~Z_{\odot}$ (using the cooling function of Sutherland \& 
Dopita 1993, which does not include effects of a UV field).
The curves are the $L_{X,bol}$-$V_c$ relationship for the simple cooling flow
models (for $\Lambda$CDM NFW haloes) described in
Sec. \ref{models}. 
The curves represent different baryonic fractions
(solid curves have $f_b=0.1$, dotted curves have $f_b=0.05$) and
abundances (thick curves: primordial abundances, thin
curves: $Z=1/3~Z_{\odot}$). }
\label{lxvc}
\end{figure}

\begin{figure}
\resizebox{\hsize}{!}{{\includegraphics{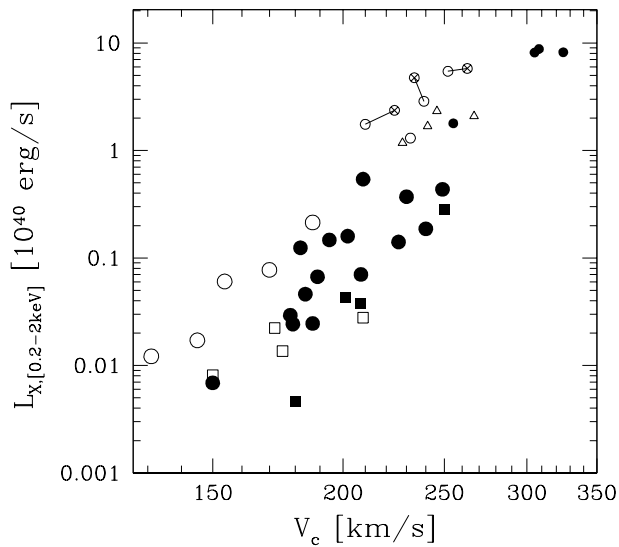}}}
\caption{0.2-2 keV band luminosity as a function of characteristic circular
speed. Symbols as in Fig. 2.}
\label{lx0.2-2vc}
\end{figure}

\section{X-ray halo emission calculation}
For each of the simulated galaxies at $z$=0 we create a catalogue of SPH gas 
particle positions, densities, temperatures and masses. 
For each catalogue a box of size (1000 kpc)$^3$ centered on 
the galaxy is
retained and all gas particles with temperatures log(T)$<4.5$ are cut away
(see Fig. \ref{figT}) since they will not contribute to
the X-ray flux (these are mainly gas particles which have cooled onto the
disc). The density and temperature of each particle is then averaged over 
itself and its five nearest neighbors using a spherical smoothing kernel
proposed by Monaghan \& Lattanzio (1985)\footnote{It is important to note that we average
over the {\it original} densities from the TreeSPH simulations.
One can show that if the cold gas particles are cut away and the
densities of the remaining gas particles are then recalculated using
the full SPH procedure the
X-ray luminosities will be underestimated due to resolution problems at the
disc--halo interface. }, 
and a volume is assigned to the particle
given its mass and density. Using the average
temperatures and densities, each SPH particle is treated as an optically
thin thermal plasma, and the associated X-ray luminosity is
calculated at the relevant position in a given photon energy
band with the {\em meka} plasma emissivity code \cite{mewe}. 
X-ray luminosities are then computed by summation over all particles in the 
volume of interest. The bolometric X-ray luminosity is calculated using the 
$0.012-12.4$ keV band.

\begin{figure}
\resizebox{\hsize}{!}{{\includegraphics{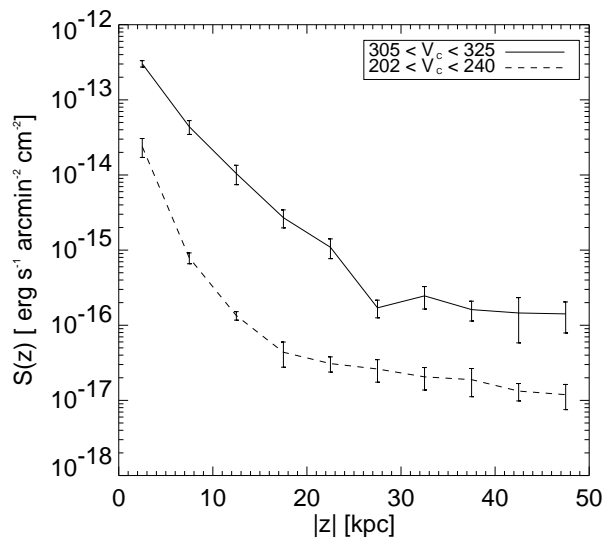}}}
\caption{Mean bolometric X-ray surface brightness
profiles perpendicular to the plane of the disc of three high $V_c$
galaxies (solid curve) and four Milky Way sized galaxies (dashed
curve). The physically most important parameters for these galaxies are $f_b$=0.10 and
primordial gas abundance.
The profiles were calculated by binning the emission in
40 kpc wide, 5 kpc high slices parallel to the disc.}
\label{surfbright}
\end{figure}  

\section{Results}
In Fig. \ref{lxvc} the total bolometric X-ray luminosities $L_{X,bol}$ of the  44 simulated disc galaxies in our sample are plotted versus their 
characteristic circular speed $V_c$, defined as the circular velocity in the
disc at $R_{2.2}= 2.2~R_d$, where $R_d$ is the disc scale length -- see
Sommer-Larsen \& Dolgov (2001) for details.
The X-ray luminosities derived from the 
simulations are up to two orders of magnitude
below values derived from simple cooling flow models which are
described in Sec. 5.1. As can be seen from the
figure $L_{X,bol}$ $\sim$
$10^{40}$ erg s$^{-1}$ for a Milky Way sized galaxy.

As expected from the simple models, the simulated galaxies display an 
$L_{X,bol}$-$V_c$ relation, but with a significant scatter. Part of this 
scatter arises from the different
conditions under which the simulations have been run (see
Sec. \ref{robustness}) and part of it is a ``real'' scatter arising
from the different geometries and cooling histories of the individual galaxy
haloes. 
This ``real'' scatter can be estimated by inspection of the large filled circles
in Fig. 2 which represent simulations run with the same, physically important
parameters (see the figure caption). These data points display a rms
dispersion of about 50\% around the mean.

There is a tendency for galaxies formed in simulations with baryon fraction
$f_b$=0.05 and primordial gas abundance to have systematically higher
$L_{X,bol}$ (by about a factor of two) than galaxies formed in similar 
simulations with $f_b$=0.10. Also, galaxies formed in simulations with gas
abundance $Z=1/3~Z_{\odot}$ tend to have systematically lower $L_{X,bol}$
than the ones with primordial gas. We discuss these trends in Sec.
\ref{models}.

Fig. 3 shows the 0.2-2 keV band X-ray luminosities of the 44 sample
disc galaxies versus $V_c$. The systematic trends mentioned above are also 
seen in this plot, in particular is the difference between the $f_b$=0.05 and
0.10 simulations (with primordial gas) even more pronounced than in
Fig. 2. We also discuss this in Sec. \ref{models}. 

Most (but not all) of the X-ray emission originates from regions of
 the hot gas halo fairly close to the disc: 95\% of the emission  typically originates
within about 20 kpc of the
disc. This is illustrated in Fig.~\ref{surfbright} where we plot the
mean surface brightness profiles perpendicular to the disc of
three high $V_c$ galaxies and four Milky Way sized galaxies.   

\section{Discussion}
\label{models}
\subsection{Comparison to simple cooling flow models}
In order to compare our results with previous work, we 
calculated a family of simple cooling flow models similar to those
considered by Benson \etal (2000). 

In these models it is assumed that the cooling occurs in a static potential
and that the gas initially was in place and traced the dark matter (DM). 
Gas is assumed to flow from the cooling radius (at which the cooling
time equals the age of the universe) to the disc (settling there as cold gas)
on a time-scale much shorter than the Hubble time. The bolometric X-ray 
luminosity can
then be approximated simply as the mass accretion rate, 
$\dot{M}_{ \rm{cool}} = 4 \pi r_{\rm{cool}}^2 \rho_{\rm{gas}}(r_{\rm{cool}})
~\dot{r}_{\rm{cool}}$, times the gravitational potential difference, so
\begin{equation}
L_{{\rm X,bol}}=\dot{M}_{{\rm{cool}}}(r_{{\rm{cool}}})\int^{{r_{\rm cool}}}_{{r_{\rm
optical}}}
\frac{V_c^2(r)}{r}{\rm d}r.
\end{equation} 

In Fig. \ref{lxvc} we show for different baryonic fractions and halo gas 
metallicities the  $L_{X,bol}$-$V_c$ relationships expected from such
simple models for haloes of the $\Lambda$CDM NFW types \cite{NFW} - the
relationships for SCDM NFW haloes are very similar. Effects of a meta-galactic
UV field were not included in these simple models --- such effects are less
significant than the effects of varying the baryon fraction and the
gas abundance. 

The disc galaxies resulting from our simulations follow an 
$L_{X,bol}$-$V_c$ relation 
with approximately the same slope as the relation derived from
simple cooling flow models ($L_{X,bol} \propto V_c^5$ - see Fig. 2), but 
shifted to systematically lower $L_{X,bol}$ and with a significant scatter. 
At small $V_c$ the relation looks steeper, but
this can be attributed to the systematically lower $L_{X,bol}$ of galaxies 
formed in simulations with a uniform gas abundance 
of 1/3 solar rather than primordial -- see sec. \ref{metal}.



The main reasons why the simple cooling flow models predict up to two orders
of magnitude more X-ray luminosity than the physically more realistic
simulated galaxies considered here are:
  
\begin{enumerate} 
\item The assumption that the DM potential is static and that all gas is in 
place at the beginning of the
galaxy's history, tracing the assumed static DM halo. In our
simulations the distribution of gas in the DM haloes is dynamic in the sense
that it changes due to discrete merging and interaction events,
especially in the early phases of the simulations, $z\ga$1-2.
\item The assumption that gas particles at the cooling radius instantly
cool out from a static infinite reservoir of hot gas. The gas halo
profile is continuously changing everywhere in the halo as the hot gas
cools out, not just at the cooling radius. Moreover, only the hot gas inside 
of the virial radius is available for disc formation --- the effect of
this in cases of high cooling efficiency is very significant as
discussed in section 5.3. 
\item The bulk of the cooling and X-ray emission is determined by what is 
going on fairly close to the disc  ($r \la 20$kpc) rather than what is 
happening at the cooling radius (see Fig.\ref{surfbright}).
\end{enumerate} 


\begin{figure}
\resizebox{\hsize}{!}{{\includegraphics{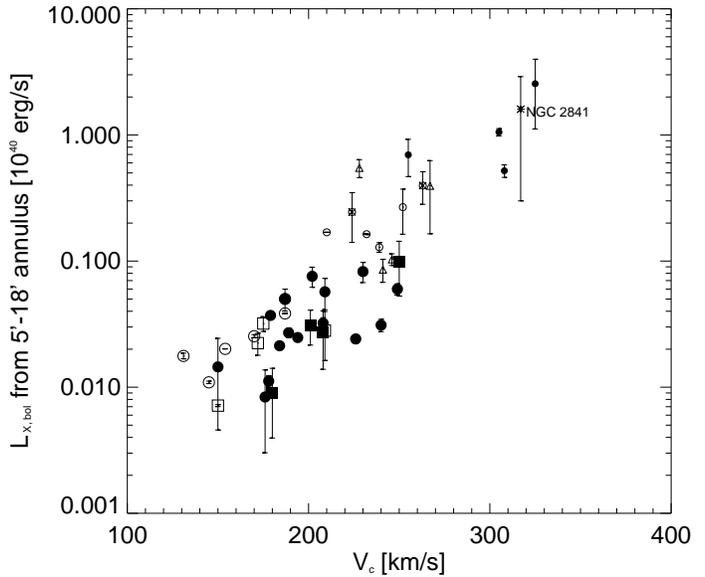}}}
\caption{Bolometric luminosity in the annulus considered by Benson \etal 
(2000) (a ring around the galaxy, with an
inner radius of 5 arcmin and an outer radius of 18 arcmin) of all
the disc galaxies in our simulation sample (assuming a distance of 13.8 Mpc).
All galaxies are viewed ``edge on'', the error bars represents the
difference in $L_X$ resulting from viewing the galaxies along the x and y 
directions, respectively. The observed data point 
(indicated by an asterisk) and its error bars are adopted from Benson \etal 
Other symbols as in Fig. 2.}
\label{lxappdist}
\end{figure}  

\subsection{Comparison to observations}
When trying to determine the amount of X-ray halo emission observationally, 
it is important not to
confuse the signature of cooling, hot gas with the emission from X-ray 
binaries in the disc, the diffuse emission of the
galaxy's disc \cite{read}, or with the X-ray emission from hot bubbles
of outflowing gas that is seen in starburst galaxies (see Pietch \etal 1997
and Strickland \& Stevens 2000).

In a recent deep Chandra
observation of a nearby edge-on spiral galaxy NGC 4631 soft X-ray
emission extending up to 7 kpc from the disc was observed
\cite{wang}. However, this disc is disturbed due to interaction and
the off--disc emission was attributed to processes in the disc.

Benson \etal (2000) attempted to detect diffuse halo emission in 
deep ROSAT observations of the haloes of three large and nearly edge-on,
isolated and undisturbed disc
galaxies; NGC 2841, NGC 4594 and NGC 5529. This study was
limited by the sensitivity, and poor point source removal
possible with ROSAT, but despite this the authors were able to
place  limits on the X-ray emission in an annulus around the galaxies
(with an inner radius of 5~{arcmin} and an outer radius of
18~{arcmin}). Of the three galaxies considered
the strongest constraint was derived for the halo of the relatively
nearby (distance $13.8 \pm 5.2$ Mpc) massive ($V_c=317\pm2$~km
s$^{-1}$) disc galaxy NGC 2841. In the annulus around this galaxy a
bolometric luminosity of $L_X=1.6\pm1.3 \times 10^{40}$erg s$^{-1}$
was derived.         
This is much less than expected from the simple cooling-flow models.
Benson \etal (2000) predict (assuming $f_b$=0.06 and $Z$=0.3 $Z_{\sun}$) the luminosity in this annulus to be about 30 times higher;
$L_X=45\pm4.4\times 10^{40}$erg s$^{-1}$ for NFW haloes (and even more for
isothermal sphere haloes). However, the above result is in excellent agreement
with the expectation from our simulations, as illustrated in
Fig. \ref{lxappdist} where we plot the bolometric luminosity in the
considered annulus (assuming a distance of 13.8 Mpc) of all the galaxies in
our simulation sample versus $V_c$. We note that our three ``data'' points at
$V_c$=300-325 km s$^{-1}$ are for simulations with baryon fraction
$f_b$=0.10 and primordial gas abundance. For a {\it given} $V_c$ we would expect the bolometric luminosities to be about twice as much for
simulations with $f_b$=0.05 - see section 5.3.4. Such a low $f_b$ is
unlikely given determinations of the ``universal'' baryon fraction,
$f_b \approx 0.10$~$(h/0.7)^{-3/2}$, derived from galaxy clusters
\cite{EF} and
galaxy clustering \cite{per}, but in any case  the
X-ray luminosities would still be consistent with the NGC 2841 measurement. 
Moreover, the (more realistic) inclusion of some level of enrichment
in the gas and the use of the more realistic meta-galactic UV field of
Haardt \& Madau (1996) would tend to {\it lower} the X-ray luminosities.  
We also note that the total X-ray luminosities of our three galaxies
at $V_c$=300-325 km s$^{-1}$ are ``only'' a factor of 3-5 lower than
the predictions of the simple cooling flow models of Benson \etal and
ours (the latter for $f_b$=0.10 and primordial gas). Hence an
important part of the reason why our models match the observed
bolometric luminosity of the 5-18 arcmin annulus around NGC 2841 is
that most of the X-rays are emitted from the inner 20 kpc of our model
disc galaxy haloes (5 arcmin correspond to 20 kpc at a distance of
13.8 Mpc). In other words our ``geometric correction'' is considerably
larger than the one used by Benson \etal

The observational constraints on the bolometric luminosity of NGC
4594 and NGC 5529 are weaker than for NGC 2841.  
The upper limits on these are again about an order of magnitude less than 
expected
from the simple models, but in agreement with the expectation
from our simulations.

The diffuse X-ray luminosity of the
Milky Way's hot halo has recently been estimated: Pietz \etal
(1998) estimate a 0.1-2 keV luminosity of 7$\cdot$10$^{39}$erg s$^{-1}$ and
Wang (1997) a 0.5-2 keV luminosity of 3$\cdot$10$^{39}$erg s$^{-1}$. Assuming
a temperature of 0.15 keV (Georgantopoulus \etal 1996; Parmar \etal
1999) this translates into 0.2-2 keV
luminosities of 5 and 7$\cdot$10$^{39}$ erg s$^{-1}$ respectively. These
estimates (which should probably be seen as upper limits) 
are consistent with
our findings from the simulations for $V_c \simeq$ 220 km s$^{-1}$ --- see Fig. 3.

\subsection{Effects of numerical resolution and physical parameters}
\label{robustness}
The galaxies in our sample have been compiled from a number of
simulations which have been run with different
cosmological and environmental parameters. These in-homogeneities in our 
simulation sample may introduce 
some scatter in the $L_X-V_c$ diagram. On the other hand,
this allows us to investigate trends when varying the physical parameters.

In general, varying a parameter has impact on the present day X-ray
luminosity of a given galaxy if it significantly alters the ability of
the hot halo gas to cool during its life time.
In the simple cooling flow models, increasing the
cooling efficiency leads to an increase in $L_X$, while this is not
necessarily the case in more realistic simulations.  
If the cooling efficiency is increased, more gas 
has cooled out on the disc at $z=0$. This results in an increase of the
characteristic circular speed of the disc as its dynamics become more
baryon (and less DM) dominated, and usually a
{\it decrease} in the X-ray luminosity of the halo since there is less hot
gas left in the halo to cool and contribute to the X-ray emission.     

In the following we briefly discuss how the derived results depend on
the resolution of the simulation, the
presence of an external UV field, the assumed gas metallicity, the
baryon fraction $f_b$, the cosmology, the dark matter type and whether
or not star formation is incorporated in the simulations.

\subsubsection{Effects of resolution}
An important test of all numerical simulations is to check whether the
results depend on the resolution. This can be done from
Fig. \ref{lxvc} by comparing the connected symbols. These represent
the same galaxy, run with normal (open
symbols) and 8 times higher mass + 2 times higher
force resolution  (open symbols with
crosses). It can be seen that this significant increase in resolution only leads to
a very modest increase of 19$\pm$64~\% in the X-ray luminosity relative to the
mean $L_X-V_c$ relation (i.e. taking the effect of the change of $V_c$ with
resolution into account). A similar result is inferred from Fig. 3.

\subsubsection{Effects of a meta-galactic UV field}
The main effects of a hard UV photon field  is to ionize the gas, significantly
reducing its ability to cool by collisional excitation
(line-cooling) mechanisms \cite{vedel}.  
This effect is evident in
Fig. \ref{lxvc} where the set of 4 small triangles and the set of 4
small open 
circles represent the same 4 haloes, run under exactly the same conditions,
except that for the latter effects of a meta-galactic, redshift-dependent UV
field were included in the cooling/heating function. There is a
tendency for the galaxies without an external
UV-field to have lower $L_{X,bol}$ and higher $V_c$ than the galaxies 
with external UV-field: The former have a median bolometric luminosity
of 55 $\pm$ 16\% of the latter (again taking into account the change
of $V_c$).   

So in this case an increase in the 
cooling rate leads to a decrease in $L_{X,bol}$ at the present epoch since a smaller
amount of hot gas is left in the halo to produce the emission -- see
 Fig.6. Note however that in the above simulations with UV field we
used a field of the Efstathiou (1992) type which is too hard and
intense compared to the more realistic one of Haardt \& Madau (1996)
--- see also Sec. 5.3.5. Hence, the suppression in cooling efficiency and
the related increase in $L_{X,bol}$ for disc galaxy haloes formed in
simulations with a Efstathiou type UV field is
somewhat too large.

\subsubsection{Effects of gas metallicity}
\label{metal}
The effects of the metallicity of the gas on the
derived $L_{X,bol}$ can be investigated by comparing the squares in
Fig.\ref{lxvc} (which have $Z=1/3~Z_{\odot}$) with the rest of the symbols
(which have primordial abundance).
In the simple cooling flow models, an increase of the metal abundance leads to an
increase in the cooling rate and $L_{X,bol}$ (compare the thick and thin
curves in Fig.\ref{lxvc}); however this is not what we find from our
simulations.  The galaxies with $Z=1/3~Z_{\odot}$ have systematically lower $L_{X,bol}$ than the
galaxies with primordial abundance, by about a factor of 3-4 for the
same $f_b$. This
is in agreement with the argument that increasing the cooling
efficiency leads to a decrease in $L_{X,bol}$ at $z=0$.
Note that for the $Z=1/3~Z_{\odot}$ simulations we used the cooling
function of Sutherland \& Dopita (1993) which does not include the
effect of a UV field. So we can not completely disentangle the effects
of gas metal abundance versus lack of UV field on the X-ray
luminosities, but Fig.6 strongly hints that the former is the most important.
\begin{figure}
\resizebox{\hsize}{!}{{\includegraphics{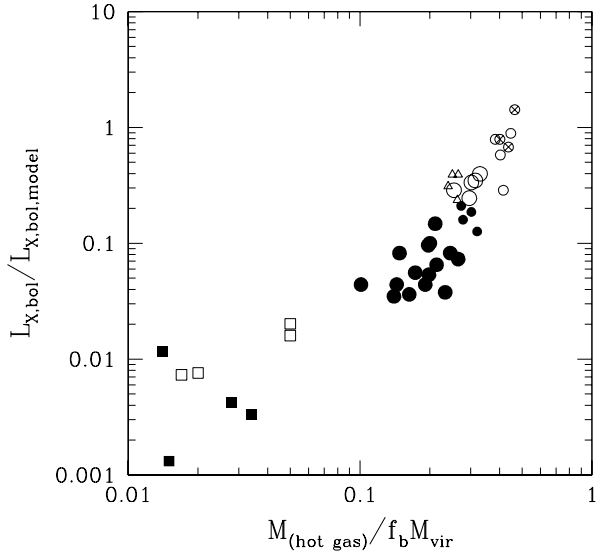}}}
\caption{Bolometric luminosity relative to the one expected for the simple
cooling flow models versus hot gas fraction - see text for details. 
Symbols as in Fig. 2.}
\label{dlxmhot}
\end{figure}

\subsubsection{Effects of baryon fraction}
Comparing in Fig.~2 $L_{X,bol}$ for simulations run with primordial gas and
with $f_b$=0.05 and 0.10, respectively, the former are more luminous on average by
about a factor of two. Yet again we see how increased cooling efficiency leads
to a decreased $L_{X,bol}$ at $z$=0.\\[1cm]
Summarizing the above results we find no statistically significant dependence
of the X-ray luminosities of the simulated galaxies on numerical resolution.
With respect to cooling efficiency there is
a general trend of higher cooling efficiency over the
course of the simulation to result in less hot gas left in the halo at $z$=0
to cool, yielding a lower X-ray luminosity. 
This is qualitatively demonstrated in Fig. 6, which
shows the bolometric luminosity relative to the one expected for the simple
cooling flow models versus the hot gas fraction 
$f_{(hot~gas)}=M_{(hot~gas)}(<r_{vir})/(f_b M_{vir}$), where 
$M_{(hot~gas)}(<r_{vir})$ is the mass of hot gas (log($T$)$>$4.5) inside of the
virial radius $r_{vir}$ and $M_{vir}$ is the total mass (baryonic + DM) inside
$r_{vir}$. 
It is seen that only for hot gas fractions $f_{(hot~gas)}\ga$ 0.4-0.5,
requiring a physically implausible parameter combination of
$f_b$=0.05, primordial gas and an unrealistically hard and intense UV field, can our
models match the  $L_{X,bol}$ predicted by the simple cooling flow models.

Pulsar dispersion measures can be used to place observational upper
limits on the amount of hot gas in the halo of the Milky Way. We find
from our simulations that Milky Way sized galaxies formed in 
primordial gas simulations have about 10$^9 M_{\odot}$ of hot
gas inside of
50 kpc at $z$=0 and the ones from the $Z=1/3~Z_{\odot}$ simulations about 10$^8 
M_{\odot}$. Both values are consistent with the observational upper limits
of about $2\cdot10^9M_{\sun}$ from
pulsar dispersion measures to the Magellanic Clouds and the globular cluster
M53 (Moore \& Davis 1994; Rasmussen 2000).

The difference between the $f_b$=0.05 and 0.10 cases is even more pronounced
for the 0.2-2 keV X-ray luminosities, as shown in Fig. 3. The reason is that
at a {\it given} characteristic circular speed $V_c$ the dynamics of the inner 
galaxy (where $V_c$ is determined) are more baryon dominated for $f_b$=0.10
than for $f_b$=0.05. This in turn means that the hot halo is smaller and
cooler for the $f_b$=0.10 case than for the $f_b$=0.05 case. 
This is demonstrated in Fig. 7, which shows the average temperature of the
central hot halo gas (inside of 20 $h^{-1}$kpc) for the 44 simulations. At
a given $V_c$ the temperature of the inner, hot halo is systematically shifted
to lower values for $f_b$=0.10 as compared to $f_b$=0.05. 
Hence for the
relevant, relatively low temperatures ($T \la 0.3$ keV, or equivalently,
$V_c \la$ 300 km s$^{-1}$) less of the emitted radiation has energies above 0.2 keV 
for the former than for the latter case. 
On the issue of baryon dominance of the inner galaxy dynamics note that
 for a {\it
given} DM 
halo, $f_b$=0.10 (as compared to $f_b$=0.05) leads to a larger $V_c$ and a 
smaller d$v_c(R)$/d$R$ in the outer parts of the disc, where $v_c(R)$
is the rotation curve -- this is in line with the findings of Persic, Salucci \& Stel 
(1996) on the basis of a large observational sample of disc galaxy rotation
curves.

Finally, as mentioned in Sec.~5.2, the temperature of the Milky
Way's inner halo is about 0.15 keV corresponding to 
1.5-2$\cdot10^6$~K in agreement with Fig.7.
\begin{figure}
\resizebox{\hsize}{!}{{\includegraphics{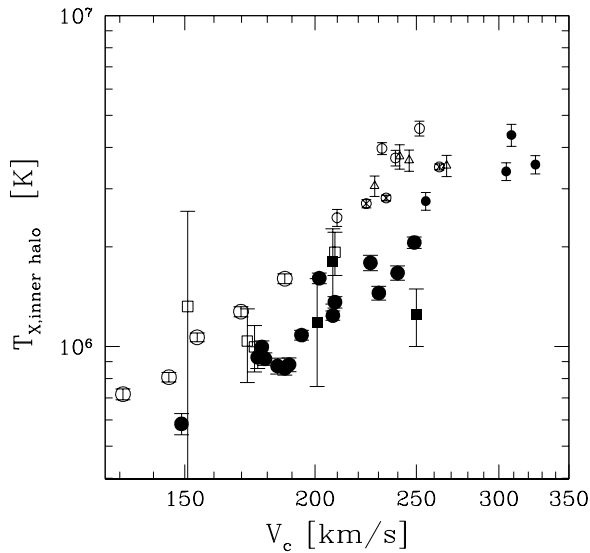}}}
\caption{Average inner halo hot gas temperature versus characteristic circular
speed - see text for details. Symbols as in Fig. 2.}
\label{Tx}
\end{figure}

\subsubsection{Effects of dark matter type, cosmology and star formation}

We do not find any dependence on the DM type, i.e. whether the 
simulations are of the WDM or CDM~+~feedback type. Neither do we find any 
indications of systematic trends with cosmology (SCDM/WDM versus 
$\Lambda$CDM/WDM) although more overlap in figures 2 and 3 between the two 
cosmologies would have been desirable (the reason for this lack of
overlap is that rather different cosmological volumes were sampled in the two
cosmologies --- the box size was $\sim 40 h^{-1}$Mpc for the
$\Omega_M$=1 cosmology and $10 h^{-1}$Mpc for the $\Lambda$-cosmology). Note also that in Fig. 6 the results
for the different cosmologies fall along the same continuous sequence. The
reason why the disc galaxies formed in primordial gas simulations for the
$\Omega_m=1$ cosmology tend to have slightly higher relative bolometric
luminosity {\it and} $f_{(hot~gas)}$ than the $\Lambda$-cosmology ones is most
likely due to the two different models of the meta-galactic UV field used:
For the former we used the one suggested by Efstathiou (1992) (see
Vedel \etal 1994), whereas for the latter we used the more realistic one
from Haardt \& Madau (1996). The former has $z_{reionization} = \infty$ and
is considerably harder and, at $z \ga 2$, more intense than the latter 
which has $z_{reionization} \simeq 6$.

Finally, we do not find any dependence on whether star formation is
included or not in the simulations.


\subsection{Mass accretion rates}
In Sommer-Larsen \etal (2002) disc gas accretion rates due to cooling-out of
hot gas are determined for the $\Lambda$CDM + feedback model disc
galaxies considered there (a subset of the sample considered here obtained in simulations with $f_b$=0.10, primordial
gas composition and the Haardt \& Madau UV field). For Milky-Way sized
galaxies accretion rates in the range 0.3-0.6 $h^{-1}$M$_{\odot}$yr$^{-1}$ ($h$=0.65)
are found at $z$=0. 
One can show that the rate at which hot halo gas cools out is proportional 
to $L_{{\rm X,bol}} \cdot <\frac{1}{T}>$, where $<\frac{1}{T}>$ is the 
emissivity 
weighted inverse temperature of the hot gas. As mentioned previously we find
in the current work that $L_{{\rm X,bol}}$ is proportional to about the fifth
power of $V_c$, so as $T \propto V_c^2$ we would expect $\dot{M} \propto 
V_c^3$. This is indeed found by Sommer-Larsen \etal (2002) and is sensible
since the $I$-band (and hence approximately mass) Tully-Fisher relation has
a logarithmic slope of about three (Giovanelli \etal 1997). 

Given the trends of $L_X$ with various environmental parameters discussed in 
section 5.3 we would expect galaxies formed in simulations with $f_b$=0.05
and primordial abundance to have about twice as large mass accretion rates,
whereas galaxies formed in $Z=1/3~Z_{\odot}$ simulations will have accretion
rates 3-4 times lower than the similar primordial gas ones. Hence we expect
at $z$=0 a fairly strong trend of the ratio of present to average
past accretion rate {\it decreasing}  with {\it increasing} cooling efficiency. Indeed  
Sommer-Larsen \etal (2002) find for their $\Lambda$CDM simulations accretion 
rates at $z$=1 which are an order of magnitude larger than the ones at $z$=0,
so disc galaxies may have been considerably more X-ray luminous in the past 
than they are today.

A detailed analysis of mass accretion rates and high-$z$ X-ray properties of
our sample of simulated disc galaxies will be presented in a forthcoming 
paper.

\subsection{Future X-ray observational tests}
From the predicted X-ray surface brightness profiles (Fig.4)
and the halo temperatures (Fig.7) we have estimated the feasibility for
detecting halo emission with XMM-Newton and Chandra using
the most recent instrument responses.
In order to avoid confusion with X-ray emission originating in the
disc we aim at detecting halo emission from nearly edge-on disc
galaxies at (vertical) disc heights of 10-15 kpc. 
Count rates for a 5 kpc high and 40 kpc wide slice (parallel to
the disc) at such disc
heights were calculated assuming a column density
of absorbing neutral hydrogen in the disc of the Milky Way of $n_H=2.5\cdot
10^{20}$~cm$^{-2}$ (corresponding to the typical value for 
galactic latitudes of $|b|$$\sim$60$^{\circ}$).
For galaxies with circular speeds in excess of 300~km~s$^{-1}$ and distances $d$$\la$50 Mpc XMM-Newton, should be able to obtain a $5\sigma$
detection at such disc heights in a 10 ksec exposure.
Although Chandra has a smaller collecting area than XMM-Newton
the Chandra background is generally lower and its superior spatial resolution
allows for more efficient removal of contaminating point sources.
We thus expect that only slightly longer exposures are required for Chandra detection of halo emission than for XMM-Newton.

However, the curve in Fig.4 for the $V_c>300$~km~s$^{-1}$ galaxies represents an
optimistic case since the underlying simulations were run with primordial
abundances and a strong external UV-field, both increasing the present
day X-ray luminosity. As mentioned in sections 5.3.2 and 5.3.3, in more realistic
simulations, including metals and a weaker external UV-field,
the halo flux is lower by a factor of about 3.
In this case, XMM-Newton as well as Chandra should still obtain
a $5\sigma$ detection for $V_c>300$~km~s$^{-1}$ galaxies within 25 Mpc in
about 25 ksecs.

For Milky Way sized galaxies, due to their
much lower surface brightness (e.g. Fig.4) and lower halo temperature
(the latter making these more sensitive to absorption) the predicted 
XMM-Newton and Chandra halo count rates are 
two orders of magnitude lower than for the $V_c>300$~km~s$^{-1}$ galaxies. Detection of
X-ray haloes for Milky Way sized galaxies at vertical disc heights of
10-15 kpc will thus have to await
future X-ray observatories with much larger collecting areas (Constellation-X
and XEUS).

\section{Conclusion}

 We have presented X-ray properties of the hot gas haloes
of disc galaxies derived from a large sample of physically realistic
gravity/hydro simulations of galaxy formation and evolution.

The simulated galaxies follow an L$_{X,bol}$-$V_c$  relation
with approximately the same slope as expected from simple cooling flow
models ($L_{X,bol}\propto V_c^5$), but shifted to lower 
$L_{X,bol}$, and with a significant
scatter (approximately a 50\% rms dispersion for a given choice of
physical parameters).

The total bolometric X-ray luminosities of the disc galaxy haloes are
 up to two orders of magnitude less than predicted by simple cooling
 flow models.
Hence, contrary to the order of magnitude discrepancy between simple cooling flow models and observations 
found by Benson \etal (2000), 
our models are in agreement with  ROSAT observations of
 the three massive highly inclined spirals NGC 2841, NGC 4594 and NGC
 5529. 
Furthermore, we find that our models are
 consistent with recent estimates of the diffuse 0.2-2 keV X-ray
 luminosity of the Milky Way, and also that the amount of hot gas in
 the haloes of our simulated, Milky Way sized disc galaxies is consistent with upper limits from pulsar
 dispersion measures toward the Magellanic Clouds and the globular
 cluster M53.

In contrast to what is predicted by simple cooling flow models, we
find that {\it increasing} the cooling efficiency  of the halo gas leads to
a {\it decrease} in the present day $L_X$.
The reason for this is that increasing the cooling efficiency over
the course of a simulation results in less hot gas in the halo at $z$=0
to cool (because the total amount of gas available at any given time
is always limited to the gas inside of the virial radius). This in turn leads to lower  present day accretion rates and
lower $L_{X,bol}$.
The two most important physical parameters controlling the
X-ray luminosities are the baryon fraction and the gas abundance:
for a {\it given} characteristic circular speed $V_c$, 
increasing the baryon fraction from $f_b$=0.05 to $f_b$=0.1
decreases $L_{X,bol}$ by a factor of about two, and similarly
increasing the gas abundance from primordial to $Z=1/3~Z_{\odot}$
results in a decrease in  $L_{X,bol}$ of a factor 3-4.

Concerning the spatial distribution of X-ray emission in the hot gas
haloes we find that this is centrally concentrated: about 95\% of the
emission originates within the inner $r\la$20 kpc.  

For their $\Lambda$CDM, $f_b$=0.1 and primordial gas abundance simulations Sommer-Larsen \etal (2002) find present
day mass accretion rates of 0.5-1 $M_{\sun}$yr$^{-1}$ for Milky Way
sized disc galaxies, compatible with observational limits (J. Silk,
private communication). As the mass accretion rate effectively is
proportional to $L_{X,bol}$ one would expect even lower accretion
rates for the more realistic case of some level of metal enrichment of
the hot gas. Furthermore, Sommer-Larsen \etal (2002) find mass accretion rates
which are an order of magnitude larger at redshift $z\sim$1. Hence it is quite
likely that disc galaxies were considerably X-ray brighter in the
past.

Forthcoming XMM-Newton and Chandra observations of massive, nearby, edge-on
disc galaxies will provide constraints on the present
models, in particular the main physical parameters: the baryon
fraction and the hot gas abundance. Future observations to redshifts
$z\ga$1 may be used  to constrain the models further.



\section*{Acknowledgments}
This project was supported by the Danish Natural Science Research 
Council (SNF) and by Danmarks Grundforskningsfond through its support
for the establishment of the Theoretical Astrophysics Center.  
We thank R. Bower, M. G\"otz, B. Moore, L. Portinari and J. Silk for useful discussions.

\label{lastpage}

\end{document}